\documentclass[12pt,preprint]{aastex}
\usepackage{color}

\newcommand\etal{{\it et al.}}

\newcommand\ie{{\it i.e.,~}}
\newcommand\K{~{\rm K}}
\newcommand\kms{~{\rm km~s^{-1}}}
\newcommand\yr{~{\rm yr}}
\newcommand\Myr{~{\rm M yr}}
\newcommand\pc{~{\rm pc}}
\newcommand\kpc{~{\rm kpc}}
\newcommand\Msun{~{M}_{\sun}}

\slugcomment{draft of \today}
\shorttitle{Origin of GW 123.4--1.5}
\shortauthors{Baek et al.}

\begin{document}

\title{How was the mushroom-shaped GW 123.4--1.5 formed in the Galactic disk?}
\author{Chang Hyun Baek\altaffilmark{1,2}, Takahiro Kudoh\altaffilmark{2,3},
\and Kohji Tomisaka\altaffilmark{2,3}}
\altaffiltext{1}{Astrophysical Research Center for the Structure and Evolution of the Cosmos (ARCSEC),
Sejong University, Seoul 143-747, Korea; chbaek@pusan.ac.kr}
\altaffiltext{2}{National Astronomical Observatory of Japan, 2-21-1, Osawa, Mitaka, Tokyo, 181-8588 Japan}
\altaffiltext{3}{Department of Astronomical Science,
 The Graduate University for Advanced Studies (SOKENDAI)}

\begin{abstract}
The unusual mushroom-shaped HI cloud, GW 123.4--1.5, is hundreds of parsecs in size 
but does not show any correlations to HI shells or chimney structures.
To investigate the origin and velocity structure of GW 123.4--1.5,
we perform three-dimensional hydrodynamical simulations of the collision of a high-velocity cloud with the Galactic disk.
We also perform a parameter study of the density, radius, and incident angle of the impact cloud.
The numerical experiments indicate that
we reproduce the mushroom-shaped structure which resembles GW 123.4--1.5
in shape, size, position-velocity across the cap of the mushroom, and the density ratio between the mushroom and surrounding gas.
GW 123.4--1.5 is expected to be formed 
by the almost head-on collision of a HVC with velocity $\sim 100 \kms$ and mass $\sim 10^5 \Msun$ about $5 \times 10^7 \yr$ ago.
A mushroom-shaped structure like GW 123.4--1.5 must be infrequent on the Galactic plane, 
because 
the head-on collision which explains the mushroom structure seems rare for observed HVCs.
HVC-disk collision explains not only the origin of the mushroom-shaped structure 
but also the formation of a variety of structures like shells, loops, and vertical structures in our Galaxy.
\end{abstract}


\section{Introduction}
GW 123.4--1.5 revealed by the Canadian Galactic Plane Survey \citep{Tay03}
is a mushroom-shaped HI structure composed of a stem and a cap 
and is dissimilar to any common shells or chimneys \citep{eng2000}.
Using the kinetic distance of the Galactic rotation model,
\citet{eng2000} derived properties of the mushroom cloud
such as projected length, mass, velocity, 
mean excess column-density, and kinetic energy.
The mushroom cloud extends a few hundred parsecs ($\sim 350 \pc$) at the assumed distance of $3.8 \pm 1.2 \kpc$
and has an HI mass of $\sim 10^5 \Msun$,  a mean excess column-density of $N_{\rm H} \simeq 6 \times 10^{20} {~{\rm cm^{-2}}}$,
and a kinetic energy of $\sim 2 \times 10^{50}$ ergs.
As a model for the origin of the mushroom-shaped cloud GW 123.4--1.5,
the gas buoyancy model \citep[]{eng2000,dm2001} 
and the cloud collision model \citep[hereafter KB2004]{kb2004} were proposed 
and numerical simulations were executed, respectively.
\citet{eng2000} carried out preliminary two-dimensional hydrodynamical simulations
and showed that the origin of the mushroom-shaped GW 123.4--1.5 is the rise of buoyant gas from
the explosion of a supernova above the Galactic midplane.
Using three-dimensional hydrodynamical simulations of large-scale modeling
of the interstellar gas in the galactic disk \citep{da2000}, 
\citet{dm2001} showed that the mushroom-shaped cloud can result from the buoyant rise of bubbles out of pools of hot gas
created from a supernova or supernovae.
However there is some difficulty in interpreting the origin of GW 123.4--1.5 
by the buoyancy of a hot gas created by a single or multiple supernova explosions,
because there are no observations of soft X-ray emission and IRAS sources
near the base of GW 123.4--1.5.
On the other hand, the collision of a high-velocity cloud (HVC) with the Galaxy, 
which was studied by \citet{Ten86} for the first time,
is not restricted to the regions of active star formation \citep{Ten87}.
KB2004 performed two-dimensional hydrodynamical simulations of the collision of a cloud with the Galactic disk
and proposed that GW 123.4--1.5 is created by the impact of an intermediate-velocity cloud (IVC)\footnotemark
\footnotetext{Historically, the clouds with $|v_{lsr}| \geq 70 \kms$ are classified into HVCs and those with $ < 70 \kms$
are called intermediate-velocity clouds (IVCs).} 
with the Galactic disk.
Also they showed that the velocity structure across the cap of the mushroom-shaped structure in their simulation
is consistent with that of GW 123.4--1.5.
In the buoyant models, the velocity characteristic of the cap is not obvious.
Unlike the gas buoyancy model,
cloud-disk collision can easily explain the formation of GW 123.4--1.5 
without inference for X-ray emissions.
 
In this study we extend the two-dimensional hydrodynamical simulation of KB2004 
to three dimensions, 
in order to investigate the origin, dynamics, and velocity structure 
of a mushroom-shaped structure.
Through numerical simulations and a parameter study, 
we can reproduce the size, shape, and position-velocity of the mushroom-shaped structure
which resemble those of GW 123.4--1.5.
Consequently, we can infer (1) how GW 123.4--1.5 was formed in the Galactic disk and
(2) why a mushroom-shaped structure is so rare on the Galactic plane.
In the next section we describe our models and numerical method. 
Simulation results are presented in \S III, 
followed by a summary and discussion in \S IV.

\section{Numerical Simulations}
We simply suppose an adiabatic gas as in KB2004. 
As a galactic disk and halo model,
we consider a hydrostatic equilibrium of the interstellar isothermal gas
determined by the gravity and temperature.
In this study, we use the Cartesian coordinates $(x, y, z)$.
The galactic disk is located at the centered $x-y$ plane of a simulation box, 
and the $z$-axis is perpendicular to the galactic disk.
Near the galactic midplane,
density distribution of the interstellar isothermal gas 
under hydrostatic equilibrium is well approximated as
\begin{equation}
\rho(z)=\rho_0 \exp{[-(z/H_0)^2]},
\label{eq:hdst}
\end{equation}
where $\rho_0$ is the density at the galactic midplane. 
The scale length is
\begin{equation}
H_0 = \sqrt{2/\gamma \alpha} c_{s0},
\end{equation}
where $\gamma$ is the specific heat ratio $5/3$
and $c_{s0}$ is the sound speed at the midplane.
The observed vertical gravitational field is approximated as
$g_z = -\alpha z$, where $\alpha \simeq 2/3 \times 10^{-29} \rm s^{-2}$
for the solar neighborhood of the Galaxy \citep[]{sp78,ba84,Ten87}.
We assume an external gravitational field and temperature distribution of interstellar gas
at the initial stage as
\begin{equation}
g_z = -\alpha H_g \tanh(z/H_g),
\end{equation}
\begin{equation}
T(z) = T_0 + 0.5(T_1 -T_0 ) \left[ 1+\tanh \left( {{|z|-z_t}\over{z_d}}\right) \right],
\end{equation}
where $H_{g}$ is $1.5H_0$ and $T_0$  and $T_1$ represent the temperatures of the disk and the halo.
We assume $T_{0}=10^4 \rm \K$, $T_{1} =10^5 \rm \K$, $z_{t} =0.5H_{0}$, and $z_{d} =0.1H_{0}$
as those in the previous two-dimensional simulations (KB2004).
As an initial condition for the Galactic disk and halo, 
we assume a hydrostatic equilibrium in the gas, which is determined by the Galactic gravity of equation (3) and the temperature distribution of equation (4).
The density distribution of the hydrostatic equilibrium is nearly equal to
equation (\ref{eq:hdst}) for $z \la z_t$, and it tends to be
$\rho (z) \propto \exp (-z/H_1)$ for $z \ga H_g$ 
where $H_1= (T_1 / T_0 )  (\gamma c_{s0}^2 /\alpha H_g)$.
In addition to this hydrostatic equilibrium gas disk,
we assume a cloud with density $\rho_c$ and velocity ${\bf v}_{c}\equiv (v_{xc},v_{yc},v_{zc})$.
Using three-dimensional velocity $v_{c0}\equiv |{\bf v}_c|$ and the incident angle $\theta_i$,
they are assumed as follows:
\begin{equation}
\footnotemark \rho_{c} = \rho_{c0} \left\{ 1-0.5 \left[ 1+\tanh \left( {r_{c} -r_{wc} \over {r_d}} \right) \right] \right\},
\end{equation}
\footnotetext{In eq. (5) of KB2004, `$1+\tanh \left({r_{c}(R,z) -r_{wc} \over {r_d}}\right)$' 
was accidentally mistyped as `$\tanh \left({r_{c}(R,z) -r_{wc} \over {r_d}}\right)$'.
The same mistake is in eq. (6) of KB2004.}
\begin{equation}
v_{xc} = \sin \theta_i \times v_{c0} \left\{ 1-0.5 \left[ 1+\tanh \left( {r_{c} -r_{wv} \over {r_d}} \right) \right] \right\},
\end{equation}
\begin{equation}
v_{yc} = 0,
\end{equation}
\begin{equation}
v_{zc} = \cos \theta_i \times v_{c0} \left\{ 1-0.5 \left[ 1+\tanh \left( {r_{c} -r_{wv} \over {r_d}} \right) \right] \right\},
\end{equation}
where $r_{c} = \left[(x-x_{c})^2 + (y-y_{c})^2 + (z-z_{c})^2\right]^{1/2}$ 
is the distance from the cloud center $(x_{c}, y_{c}, z_{c})$ and
$v_{c0}=100\kms$ is a velocity of the cloud.
$v_{xc}$ and $v_{zc}$ are parallel and perpendicular to the galactic disk, respectively.
We take $x_c = y_c = 0$, $z_c = -2.5 H_0$, $r_{wc} = 0.4H_0$ or $0.6H_0$, $r_d = 0.1 H_0$, and
$r_{wv} = r_{wc} + 2r_d$ for a colliding cloud.

We perform three-dimensional numerical simulations 
using an Eulerian hydrodynamics code based 
on the total variation diminishing scheme \citep{ryu93}. 
While radiative cooling and magnetic fields are likely to be important to the dynamics
of interstellar gas in the galactic disk,
we ignore them due to numerical complexity and limited computational resources.
Radiative cooling and magnetic field will be considered in an upcoming paper.
Simulations are made with $256 \times 256 \times 512$ grid zones in $x$, $y$, and $z$-directions respectively,
and the computational box size is taken $0.7\kpc \times  0.7\kpc  \times 1.4\kpc$.
A free boundary condition is imposed on all outer boundaries
of the computational box.
Simulations start at $t = 0$ and last up to $t_{\rm end}=6 t_0$.
All physical variables are expressed in units of the following normalization:
$H_0 = 140\pc$, $c_{s0} = 10 \kms$, and $t_0 = H_0/c_{s0} \simeq 1.4 \times 10^7 \yr$.

This paper presents a total of seven simulations, 
which differ in density ($\rho_c$), radii ($r_c$), and incident angles ($\theta_i$) of impact clouds.
Model parameters are summarized in Table 1.

\section{Results}
\subsection{Head-on Collision}
Firstly, we present the case of the head-on $(\theta_i =0)$ collision
to compare with the previous two-dimensional simulation (KB2004).
Figure 1 and 2 show the evolution of the density and pressure
as well as the velocity field on the $y=0$ plane in model HH,
in which the same parameters are taken as those of KB2004;
$n_0 =1 \rm cm^{-3}$, $r_{wc} = 56  (H_0/140 \pc) \pc$. 
The top-left panel depicts the initial conditions.
The cloud collides with the galactic disk at $t/t_0 \simeq 0.2$, 
and penetrates through the dense midplane at $t/t_0 \simeq 0.3$.
At $t/t_0 \simeq 0.8$ we can see that a bow shock is traveling toward the halo
($(x,z)\simeq (0,3.7H_0)$)
and a shell-like structure (density enhancement around $(x,z)\simeq(0,2.75H_0)$)
is formed by the cloud penetration (the third snapshot).
Although the shell consists of a high-density gas, 
the pressure (see Fig.\,2) is not especially high
compared with the post-bow shock gas.
This elongated shell structure originates in the tenuous halo gas and the dense disk gas. 
The inside of the elongated shell is pervaded by a gas coming from the impact cloud
($t/t_0 \simeq 0.8 - 1.0$).
In the pressure plot (Fig.\,2), two reverse shocks appear at $t/t_0\simeq 0.8$.
One is seen at the back of the cloud ($(x, z) \simeq (0,0)$) and
the other is around $(x,z)\simeq (0,2H_0)$ at the same time.
Third and fourth snapshots of Figure 2 indicate both reverse shocks bear 
the signature of the Mach disk, which is formed by a reflection of the
shock front on the $z$-axis.
On the other hand, in the galactic disk, a cavity is left
after the cloud penetration.
Three snapshots (from the fourth to the sixth) of Figure 1 indicate 
that the cavity is gradually being refilled with the galactic disk gas.
This seems to be driven by the radially inward pressure force ($t/t_0 \simeq 1.6 - 2.0$).
The reverse shock which is formed at $t/t_0\simeq 0.8$ and propagates downward 
(seen at $t/t_0 \sim 1.6$) interacts with the gas converging to the center of the cavity.
The reverse shock hits the high-density gas at the age of $t/t_0\simeq 2.0$,
just when the cavity has been refilled.
By the interaction between the shock wave and the converging gas,
the converging dense gas is shock-heated and begins to rise along the $z$-axis 
($t/t_0 \ga 2.2$: 7th and 8th snapshots).
The gas at high latitude is broken into eddies or vortices
and forms the cap of a mushroom-shaped structure ($t/t_0 \ga 2.4$).
The shock-heated rising gas contributes to filling the central part of the stem 
which is seen in the 7th and the last snapshots  ($t/t_0 \simeq 2.4 - 2.8$).
Similarly to KB2004, the gas broken into eddies at the high latitude (cap) 
and the dense gas raised by the cavity replenishment (stem)
have finally formed a mushroom-shaped structure.

Figure 3 and 4 depict the evolution of the density and pressure as well as
velocity field in model HL. 
In this model,  the cloud is less dense $n_0 =0.1 \rm cm^{-3}$
but larger $r_{wc} = 84 (H_0 /140 \pc) \pc$  than that of model HH.
Although the initial cloud masses assumed in models HH and HL are quite similar, 
the initial kinetic energy $E_k =(1/2) \int \rho {\bf v}^2 dV$ of the HVC in model HL is 40 \%  less energetic 
than that in the previous model HH.
Because the density and velocity distributions of the cloud are not assumed 
to be uniform but to have smooth distributions (eqs. (5)--(8)).
Dynamical evolution, density, and temperature distributions of the mushroom 
in model HL differ considerably from those in model HH.
After a cloud collides with the galactic disk, a shell-like structure is formed
inside the galactic disk at $t/t_0 \simeq 1.0$ (second and third snapshots).
Even if the cloud masses in both models are similar,
the shell-like structure in model HL is different from model HH;
the lateral diameter of the shell $\sim 2.0H_0$ (third and fourth snapshots of Fig.\,3)
is much larger than  that of the previous model $\sim 1.0H_0$ (third snapshot of Fig.\,1);
in contrast, the vertical size is half that of model HH.
In the third snapshot, we can see a reverse shock at $(x,z)\simeq (0,1.25H_0)$ 
moving toward the galactic plane at $t/t_0\simeq 1.2$.
This shock wave propagates and reaches the midplane at $t/t_0\simeq 1.8$ 
(fourth snapshot).
At the same time,
the cavity is refilled with the disk gas approaching from lateral directions 
pushed by the interstellar pressure ($t/t_0 \simeq 2.0 - 2.2$).
The wave interacts with the refilled dense gas and forms a high-pressure gas,
which is seen near $z\simeq 0.5H_0$ in the 5th snapshot of Figure 4. 
This promotes the cavity replenishment and lifts the galactic dense gas along
with the vortices ($t/t_0 \ga 2.2$).
In model HH, the vortex structures have been developed
before the cavity replenishment has been finished.
However, in model HL,
the vortices are developed in the process of the cavity replenishment
at the galactic plane.
The vortices, of which the central location is rising along the stem,
play a major role in forming the cap of the mushroom.
Finally, the rising gas and vortices from the disk make
a mushroom-shaped structure ($t/t_0 \simeq 4.8 - 5.6$).
The rising gas is decelerated by the gravity of the galactic disk
and eventually falls toward the disk, as seen in the last snapshot
($t/t_0 \ga 5.2$).

Figure 5 compares the density, velocity vector field (top panels),
and temperature (bottom panels) of the mushroom-shaped structure
on the $y=0$ plane in models HH (left) and HL (right).
We can see the density and temperature differences
between models HH and HL.
In the case of model HH, the density shows a hollow in the stem
which is filled with a hot gas. 
On the other hand, the stem of model HL has a high-density and low-temperature
without any indication of the hollow.
The difference comes from the formation processes of the stem.
The stem of model HH
was first formed from the gas which was originally a high-density disk and
elongated into the low-density halo as the cloud collided with the disk
and moved upward.
Later on, a high-density and low-temperature gas blew up
and occupied the inner hollow,
after the disk was replenished and collided with a reverse shock.
On the other hand, the stem of model HL is not formed
until the disk gas blows up, 
because the disk gas is not sufficiently elongated upward
when the cloud collides with the disk. 
Then, the stem of model HH has a hot hollow structure during its evolution,
but the stem of model HL is always occupied by the dense and cold gas.
This difference can also explain the reason why the mushroom 
structure of model HL appears later than that of model HH.

KB2004 showed that the cloud density should be of the same order as the disk
density to attain a mushroom-shape.
Because the case of IVC ($v_{zc}=50 \kms$) showed 
a better agreement mainly with the height of the mushroom-shaped structure,
they proposed that GW 123.4--1.5 was created
by the impact of a high-density ($n_0 = 1 {~{\rm cm^{-3}}}$)
IVC with the galactic gas disk.
Our model HL shows that the mushroom-shaped structure
can also be made by the collision of a low-density
($n_0 = 0.1 {~{\rm cm^{-3}}}$) HVC ($v_{zc}=100 \kms$)
with the galactic disk,
if the size of the cloud is sufficiently large.
Whether for HVC or for IVC,
the cloud-disk collision finally forms a mushroom-shaped structure.
Therefore we may conclude that cloud collision with the Galactic disk is a promising model
for the mushroom-shaped GW 123.4--1.5.

\subsection{Oblique Collision}

In this section, we explore oblique collisions
with different incident angles (see Table 1).
As models of oblique collision,
we assume the same cloud in model HL
and change only the incident angle of the cloud.
Figure 6 and 7 show the column-density and initial velocity (white arrows) of a cloud on the $x-z$ plane and the $y-z$ plane, respectively.
We integrate the density along the $y$-axis (Fig.\,6) and $x$-axis (Fig.\,7)
to get a column-density at $t/t_0=2.4$ (middle panels)
and $t/t_0=4.8$ (bottom panels)
for four models with different incident angles 
($\theta_i$ is an angle between the vertical axis of the galactic disk
and the initial velocity vector of a cloud), 
$5 \arcdeg$ (model O5),
$10 \arcdeg$ (model O10), 
$20 \arcdeg$ (model O20),
and $30 \arcdeg$ (model O30), respectively.
The time evolution of model O3 ($\theta_i = 3 \arcdeg$) is shown
in Figure 8 by the snapshot of column-density distribution.
Figure 6 shows that a variety of structures
such as mushroom-shape (bottom panel of model O5),
shell/loop-like (middle panels of models O5-O30),
and vertical structures (bottom panels of models O5-O20) 
are effectively formed by cloud-disk collisions.
After the cloud-disk collision, 
a variety of structures are quickly created by a subsequent process
of cavity replenishment, gas rise, and eventual fall back
due to the gravity of the galactic disk.
The dynamical evolutions of oblique collisions are roughly similar
except for the final structures.
The resultant structures formed by cloud-disk collision 
are greatly affected by the incident angle.
When we observe the cloud-disk collision on the $y-z$ plane (Fig.\,7),
the projected structure (column-density integrating along the $x$-axis) evolutions in each model look very similar
besides resultant structures (bottom panels).
First, a shell and cavity are created, as seen at $t/t_0=2.4$ (middle panels),
then change to vertical structures which penetrate the galactic disk vertically.
Like the resultant structures observed on the $x-z$ plane (bottom panels of Fig.\,6),
the vertical structures projected at $x=+\infty$
are mainly influenced by the incident angle at the $x-z$ plane.
From this parameter test, we can comprehend that 
cloud-disk collision of the large incident angle (model O10, O20, and O30)
can hardly create a mushroom structure.
Therefore if we can observe a mushroom structure along the $x$-direction, 
the incident angle at the $x-z$ plane must be small $\theta_i \lesssim 5\arcdeg$.

Obliquely colliding clouds have initial velocity $v_{xc}$ and $v_{zc}$
given by equations (6) and (8). 
The initial trajectory of the colliding clouds is represented
by white arrows in the middle ($t/t_0=2.4$)
and bottom panels ($t/t_0=4.8$) of Figure 6.
At first glance, the resultant structures in the middle panels 
(especially for $\theta_i \geq 10 \arcdeg$)
seem to be formed by the collision of a cloud moving from right to left,
contrarily to the real situation.
As a consequence of the cloud collision,
a structure expanding in $z < 0$ region is formed
as well as the upwardly expanding main component.
This countersplash, which is named for the structure seen in $z < 0$,
seems to be a wake of the colliding cloud.
If we misidentify this splash as a wake,
we may consider the cloud moved from right to left.

In a preliminary parameter test (we do not report the simulations in this paper), 
we considered several clouds which have a variety of velocity ($\sim 30 - 200 \kms$) 
and mass ($\sim 10^4 - 10^6 \Msun$). 
When the kinetic energy of the cloud is larger than $\gtrsim 10^{53}$ ergs, 
the structures after the cloud-disk collision are $> 700 \pc$ vertical structures.
We did not explore these structures, because their size is much larger than the HI mushroom structure.
On the other hand if the kinetic energy of the cloud is smaller than  $\la 5 \times 10^{51}$ ergs (\ie $\sim 30 \kms$), 
the colliding cloud does not penetrate the disk such as described in the previous two-dimensional simulation (KB2004)
or does only make shell/loop-like structures near the galactic plane. 
In this study, our concern is focused on the formation of the mushroom-shaped cloud.
We chose the incident kinetic energy of the cloud as $\sim 10^{52}$ ergs.
Figure 6 clearly shows that the cloud-disk collision can hardly make a mushroom-shaped structure,
when the incident angle of a colliding cloud is larger than about $5 \arcdeg$. 
Among our simulation results, the model O3 ($\theta_i =3 \arcdeg$; Fig.\,8) can describe GW 123.4-1.5 most closely. 
Then, the mushroom-shaped structure GW123.4-1.5 is likely produced by the
approximately head-on collision of the cloud with a velocity of $\sim
100 \kms$, a mass of $\sim 10^5 \Msun$, and thus a kinetic energy of
$\sim 10^{52}$ ergs.

\subsection{A Model for GW 123.4--1.5}

Figure 8 shows the evolution of column-density on the $x-z$ plane (integrating along the $y$-axis) for model O3.
We can find the mushroom-shaped structures
between $t=56 (t_0/14 \Myr) \Myr$ (at $t/t_0 = 4.0$) 
and $t=67 (t_0/14 \Myr)  \Myr$ (at $t/t_0 = 4.8$).
The height of the mushroom at the age of $t=67 (t_0/14 \Myr) \Myr$
is about $350 (H_0/140 \pc) \pc$,
 which is the same size as that of GW 123.4--1.5.
The lifetime of the mushroom structure is $\sim 10^7 \yr$ 
and the width ratio of the cap to the stem is about 3:1.
The mass of the cap is about twice that of the stem, 
and the total mass is estimated to be $\sim 10^5 \Msun$.
To obtain the mass of the mushroom-shaped cloud,
we integrate the low-temperature gas with $T \leq 0.8T_{1}$.
The mean column-density contrast between the mushroom and ambient medium
is about 2:1.
Although the mass ratio is a little smaller
than the observed one 4:1 \citep{eng2000},
most physical parameters of the mushroom-shaped structure in model O3
coincide with those of GW 123.4--1.5.

In order to explain the velocity structure of the mushroom-shaped structure from the two-dimensional simulation,
KB2004 assumed that the mushroom-shaped cloud is tilted to the line-of-sight
and calculated the line-of-sight velocity from their two dimensional simulation.
However, without any presumption,
we can reproduce the position-velocity map of the mushroom-shaped cloud
from the three-dimensional simulation.
Figure 9 shows the column-density and the position-velocity maps,
which are equivalent to the observational longitude-velocity ($l-v$) maps,
for model O3.
In order to distinguish the mushroom gas from the surroundings,
we integrate only the low-temperature gas ($T \leq 0.8T_{1}$)
along the $y$-axis (Fig.\,9a) and the $x$-axis (Fig.\,9b), respectively.
Figure 9a corresponds to the case observed from $y=-\infty$ and Figure 9b from $x=+\infty$.
The position-velocity plot of Figure 9a indicates the symmetry with respect to $v_y=0$ axis 
for both the bottom ($z=2.4 H_0$) and the upper ($z= 3.2 H_0$) regions of the cap. 
On the other hand, the position-velocity plot of Figure 9b shows that
the upper part of the cap has no significant velocity gradient, but
the bottom has two distinct lobes blue-shifted with respect to the center of the cap 
by about $5 \kms$.
This velocity gradient is very similar to the observed value of GW 123.4--1.5 \citep[Fig.\,1c of][]{eng2000}.
Observed (GW 123.4-1.5) and simulated (model O3) mushroom-shaped clouds are compared in Table 2.
Using model O3 which fits the observations best,
we can infer that GW 123.4--1.5 is created by a HVC 
with a velocity of $\sim 100 \kms$ coming toward us with $v_{lsr} \sim 5 \kms$ and a mass of $\sim 10^5 \Msun$ 
collided with the Galactic disk about $5 \times 10^7 \yr$ ago.

\section{Summary and Discussion}
We perform three-dimensional hydrodynamical simulations 
for the impact of a HVC with the Galactic disk,
in order to explore the formation of the mushroom-shaped HI cloud GW 123.4--1.5. 
The main results can be summarized as follows:
\begin{itemize}
\item A mushroom-shaped structure can be formed
not only by a high-density IVC collision with the Galactic disk (KB2004) 
but also by a low-density HVC collision.
\item The resultant structures formed by cloud-disk collision 
are greatly affected by the density and incident angle of a colliding cloud.
\item GW 123.4--1.5 is expected to be formed 
by the almost head-on collision of a HVC with velocity $\sim 100 \kms$ and mass $\sim 10^5 \Msun$ about $5 \times 10^7 \yr$ ago.
\end{itemize}

Recently, \citet{as2005} have tried to find a mushroom-shaped HI structure like GW 123.4--1.5 in the Canadian Galactic Plane Survey.
However, they did not find other mushroom shape structures.
Our simulations could explain 
why mushroom-shaped structures are so infrequent in our Galaxy.
A nearly head-on collision is needed to form a mushroom-shaped structure.
In a rotating disk, 
the head-on collision of a cloud with the Galactic plane would be rare in a general way.

HVC-disk collision can not only explain the origin of a mushroom-shaped structure 
but also the formation of a variety of structures like shells, loops and vertical structures in our Galaxy.
For example, in model O30 (see the middle panel of Fig.\,6), we can find an interesting structure, a loop
which resembles that found by NANTEN in the central molecular zone 
within $\sim 1 \kpc$ from the Galactic center \citep{Fuk2006}.
The two loops of CO emission obtained with NANTEN
have strong velocity gradients in a longitude-velocity diagram:
$\sim 80 \kms$ per $250\pc$ and $\sim 60 \kms$ per $150\pc$ in each loop \citep[see Fig.\,2 of][]{Fuk2006}.
Although the projected velocity distribution of our simulation ($\sim 20 \kms$) is somewhat smaller than that of observation ($\sim 60 - 80 \kms$),
the size of the loop, $\sim 300 (H_0/140\,{\pc}) \pc$, is similar to that of the molecular loop.
In order to explain the formation of the molecular loop, 
\citet{Fuk2006} proposed a magnetic flotation model
(due to magnetic buoyancy caused by the \citet{Park66} instability).
From our preliminary result of model O30, 
we also propose that HVC-disk collision is another possible model of formation of the molecular loop structure
at the Galactic center.
This issue will be studied in detail in an upcoming paper.

The present study as well as the previous study (KB2004)
are restricted to hydrodynamical simulations. 
When the magnetic field parallel to the Galactic disk exists \citep[]{santi2000,franco2002,nozawa2005} , 
we also expect the magnetic buoyant effect (Parker instability).
Both the impact of the cloud and the Parker instability \citep[]{santi1999,santi04,santi07k}
may create complicate structures in the Galactic disk
such as worms, loops, shells, and chimneys.
Moreover the interaction between the infalling HVCs and the magnetized disk must 
be an elementary process in the interstellar dynamics.
Using high-resolution two-dimensional MHD simulation, 
\citet{santi07} pointed out a possibility that
 the rain of compact HVCs onto the disk can maintain transonic turbulent motion in the warm phase of HI. 
In this sense, we believe that high-resolution three-dimensional MHD simulation will 
broaden our understanding of the structures formed by HVC collisions
and bring us more interesting phenomena to be important in the interstellar dynamics.
In the near future, we will perform three-dimensional magnetohydrodynamical simulations.

\acknowledgments
We thank the anonymous referee for clarifying comments.
The work of  C.H.B. was supported by the Korea Research Foundation
Grant funded by the Korean Government (KRF-2006-352-C00030).
The work of K.T. was supported by KAKENHI (17340059 and 16204012) from the Japan Society for the Promotion of Science (JSPS). 
Simulations were run using VPP5000 at the National Astronomical Observatory of Japan (NAOJ) 
and a high performance cluster built
with funding from the Korea Astronomy and Space Science
Institute (KASI) and the Astrophysical Research Center for
the Structure and Evolution of the Cosmos (ARCSEC) of the
Korea Science and Engineering Foundation (KOSEF).

\clearpage
\begin{deluxetable}{ccccccc}
\tablewidth{0pc}
\tablecaption{Model Parameters for Simulations}
\tablehead{\colhead{Model} & \colhead{$n_{c0}$ (${~{\rm cm^{-3}}}$)\tablenotemark{a}}
& \colhead{$r_{wc}$}
& \colhead{$\theta_i $(deg)\tablenotemark{b}} }
\startdata
HH  &  1.0  & 0.4$H_0$ &  0.0 \\
HL  &  0.1  & 0.6$H_0$ &  0.0 \\
O3  &  0.1  & 0.6$H_0$ &  3.0 \\
O5  &  0.1  & 0.6$H_0$ &  5.0 \\
O10 &  0.1  & 0.6$H_0$ & 10.0 \\
O20 &  0.1  & 0.6$H_0$ & 20.0 \\
O30 &  0.1  & 0.6$H_0$ & 30.0 \\
\enddata
\tablenotetext{a}{$n_{c0}$ is the number density at the center of the HVC.}
\tablenotetext{b}{$\theta_i$ is an angle between the vertical axis of the galactic disk and the initial vector of a cloud.}
\end{deluxetable}

\clearpage
\begin{deluxetable}{ccc}
\tablewidth{0pc}
\tablecaption{Comparison of observed (GW 123.4-1.5) and simulated (model O3) mushroom-shaped cloud}
\tablehead{\colhead{} & \colhead{GW 123.4-1.5\tablenotemark{a}} & \colhead{model O3}}
\startdata
height ($\pc$) & 350\tablenotemark{b} & {$350 \cdot (H_0 /140 \pc)$} \\
total mass ($10^5 \Msun$) & 1.55 & 0.76 \\
mass ratio (cap:stem) & 3.4:1.0 & 1.7:1.0 \\
$\triangle v$\tablenotemark{c} ($\kms$) & 5 & 5\\
mean column-density contrast (mushroom:ambient medium) & 4:1 & 2:1
\enddata
\tablenotetext{a}{Observational data from \citet{eng2000}. }
\tablenotetext{b}{A distance of $3.8 \kpc$ was assumed in \citet{eng2000}.}
\tablenotetext{c}{Velocity difference between the lobes of the cap and the central cap.}
\end{deluxetable}

\clearpage
\begin{figure}
\plotone{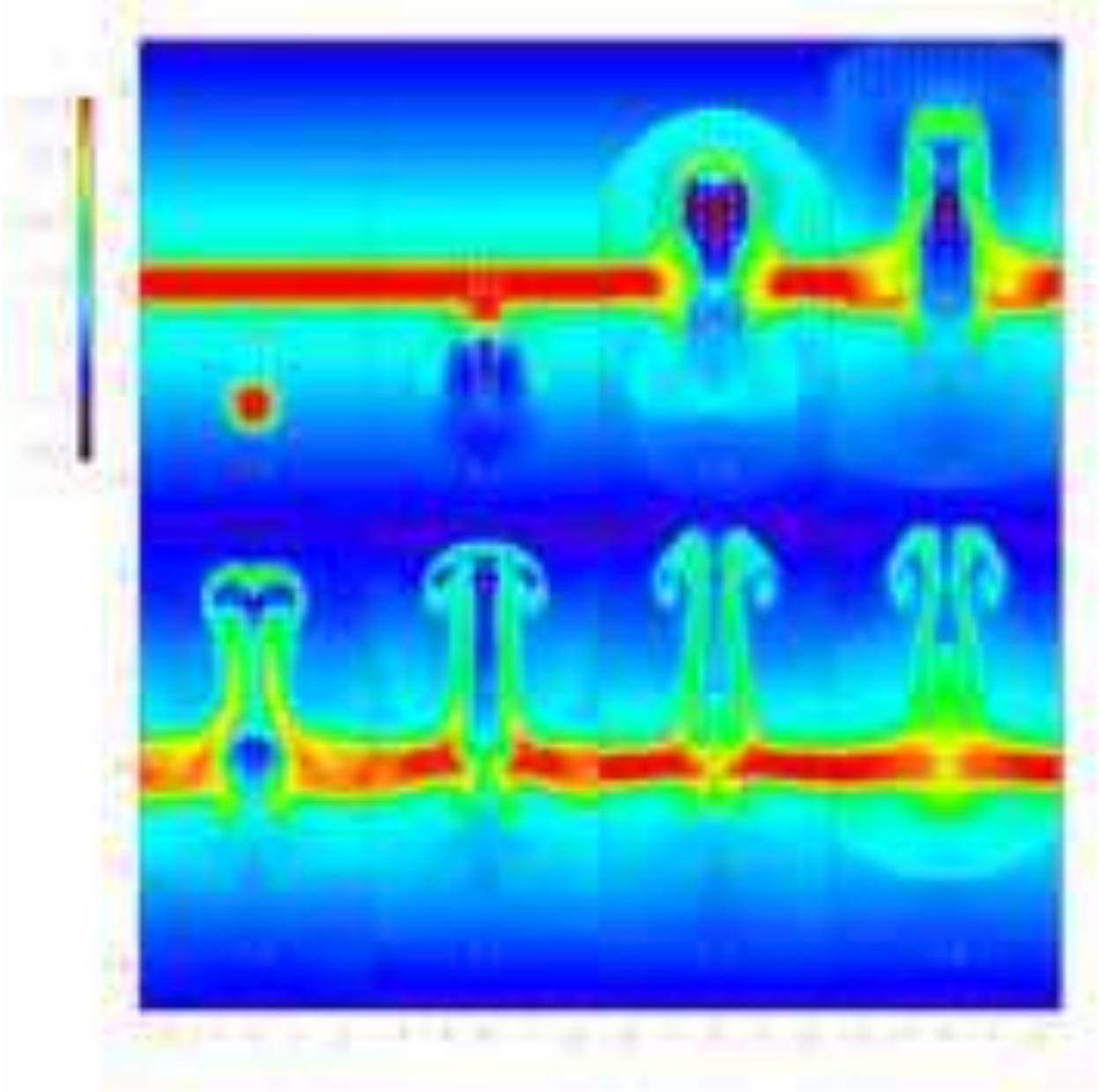}
\figcaption
{Time sequence of density distributions (in logarithmic scale)
and velocity field at the $y=0$ plane
at $t/t_0 = 0, 0.2, 0.8, 1.2$ (top:left to right),
$1.6,2.0,2.4$, and $2.8$ (bottom:left to right) for model HH.
The abscissa represents the $x$-axis and the ordinate represents the $z$-axis.
The longest arrow correspond to about $100 \kms$.}
\end{figure}

\begin{figure}
\plotone{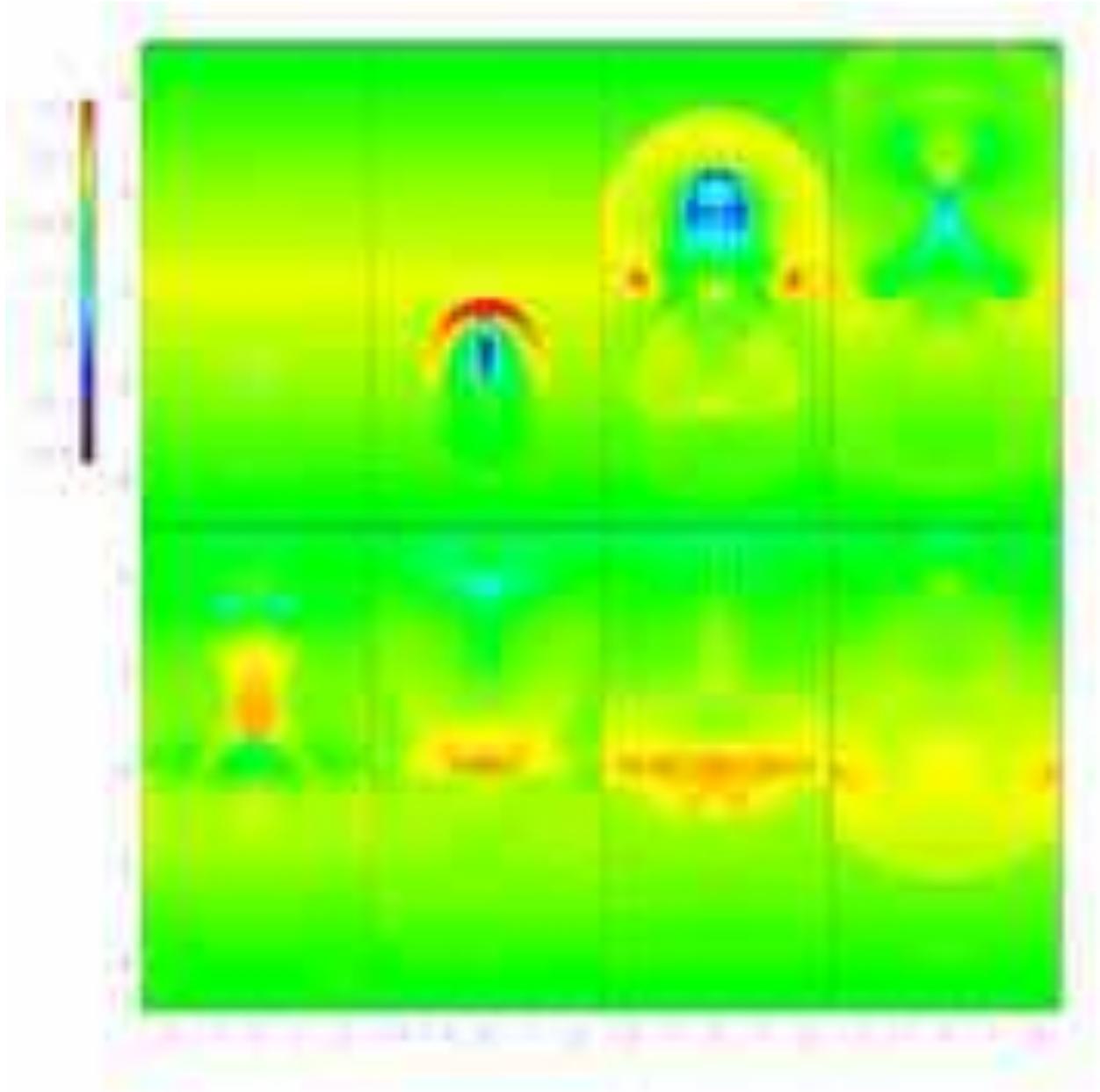}
\figcaption
{Time sequence of pressure at the $y=0$ plane 
at $t/t_0 = 0, 0.2, 0.8, 1.2$ (top:left to right),
$1.6,2.0,2.4$, and $2.8$ (bottom:left to right) for model HH.}
\end{figure}

\begin{figure}
\plotone{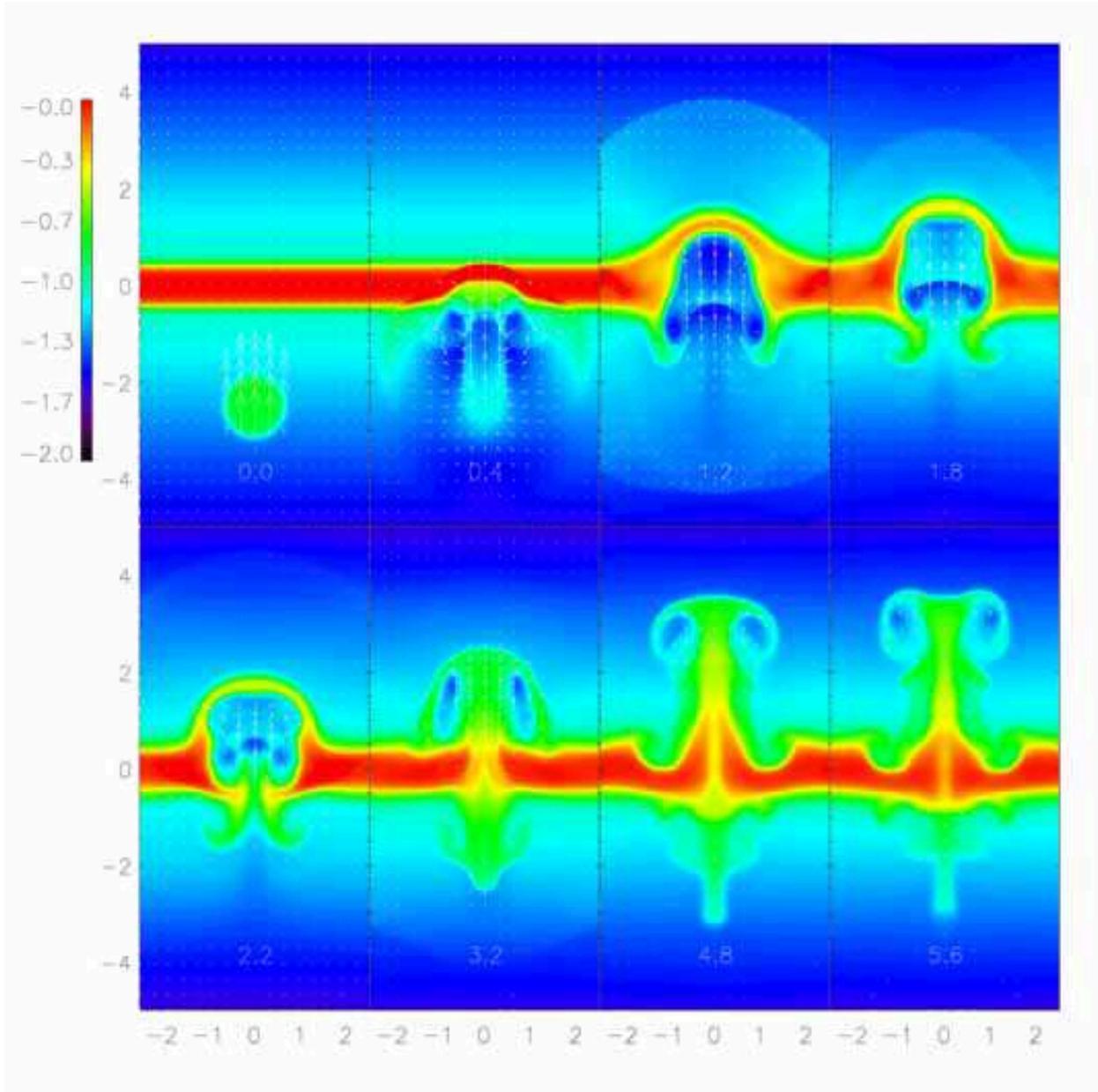}
\figcaption
{Same as Fig.\,1, but for model HL. 
Time sequence is $t/t_0 = 0, 0.4, 1.2, 1.8$ (top:left to right), 
$2.2, 3.2, 4.8$, and $5.6$ (bottom:left to right).}
\end{figure}

\begin{figure}
\plotone{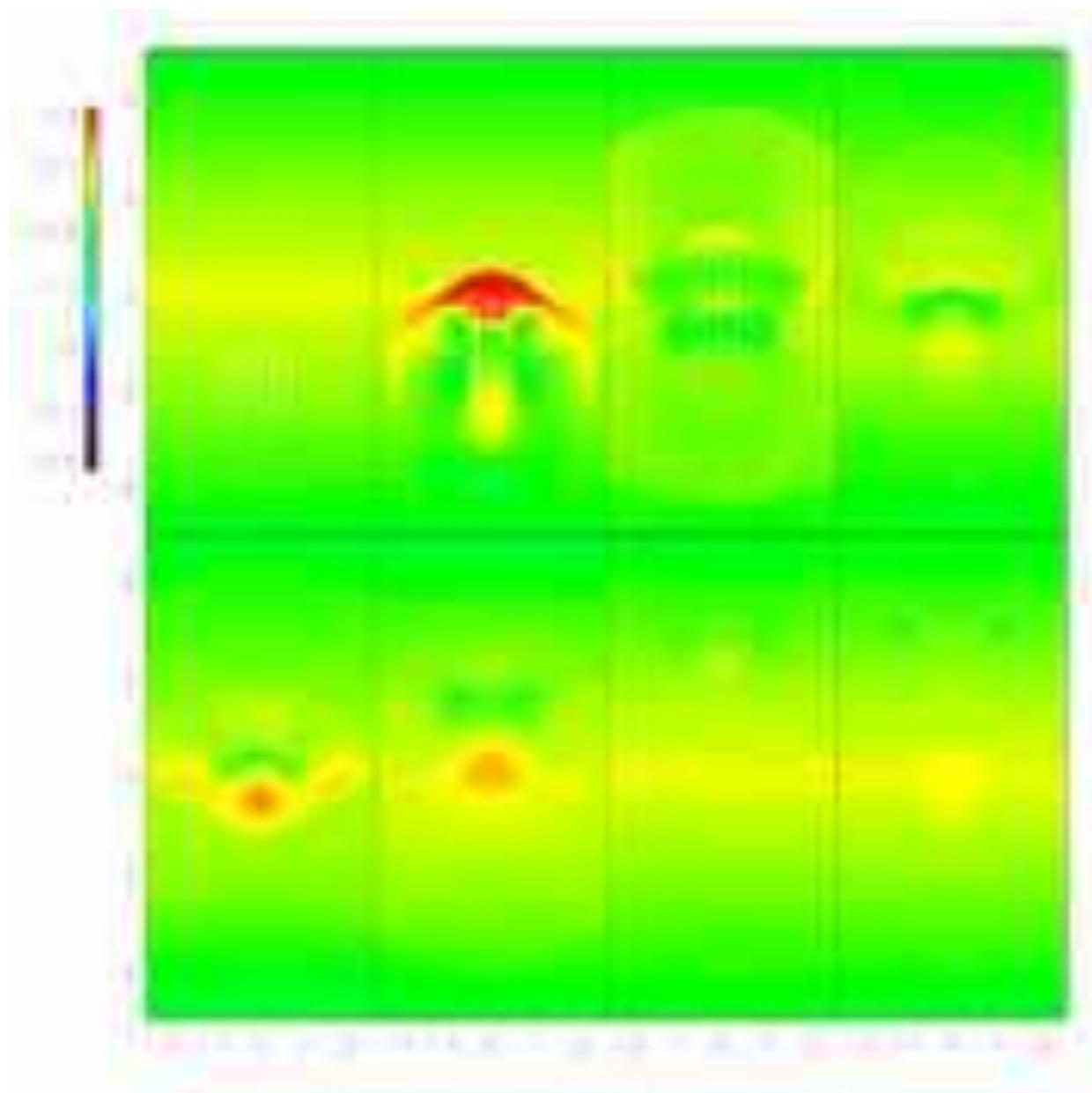}
\figcaption
{Same as Fig.\,2, but for model HL.}
\end{figure}

\begin{figure}
\plotone{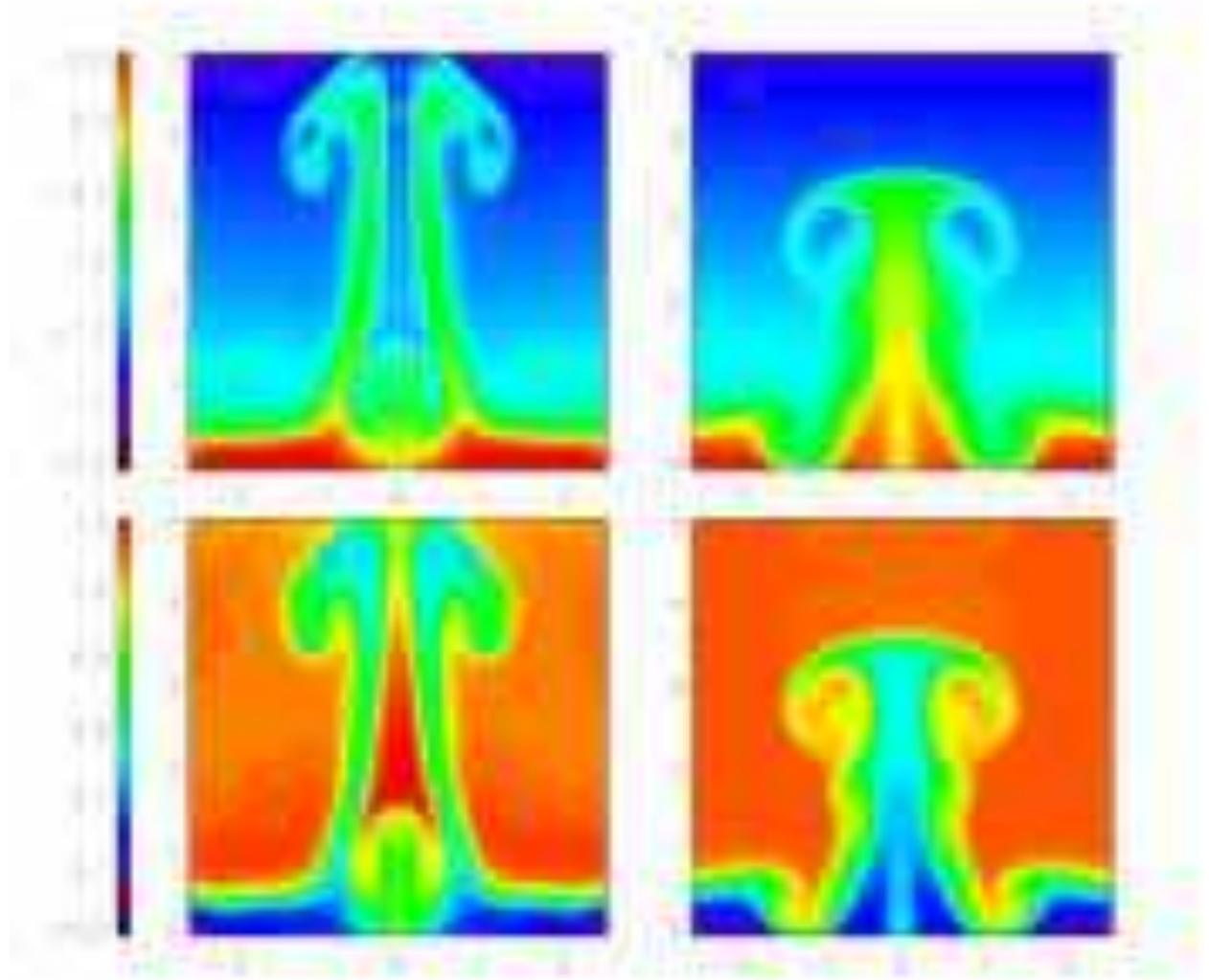}
\figcaption
{Density, velocity (top panels) 
and temperature distributions (bottom panels) on the $x-z$ plane.
(a) is for model HH at $t/t_0=2.4$ and 
(b) is for model HL at $t/t_0=4.8$.
Temperature (in logarithmic scale) is normalized in $10^4 \rm K$.}
\end{figure}

\begin{figure}
\plotone{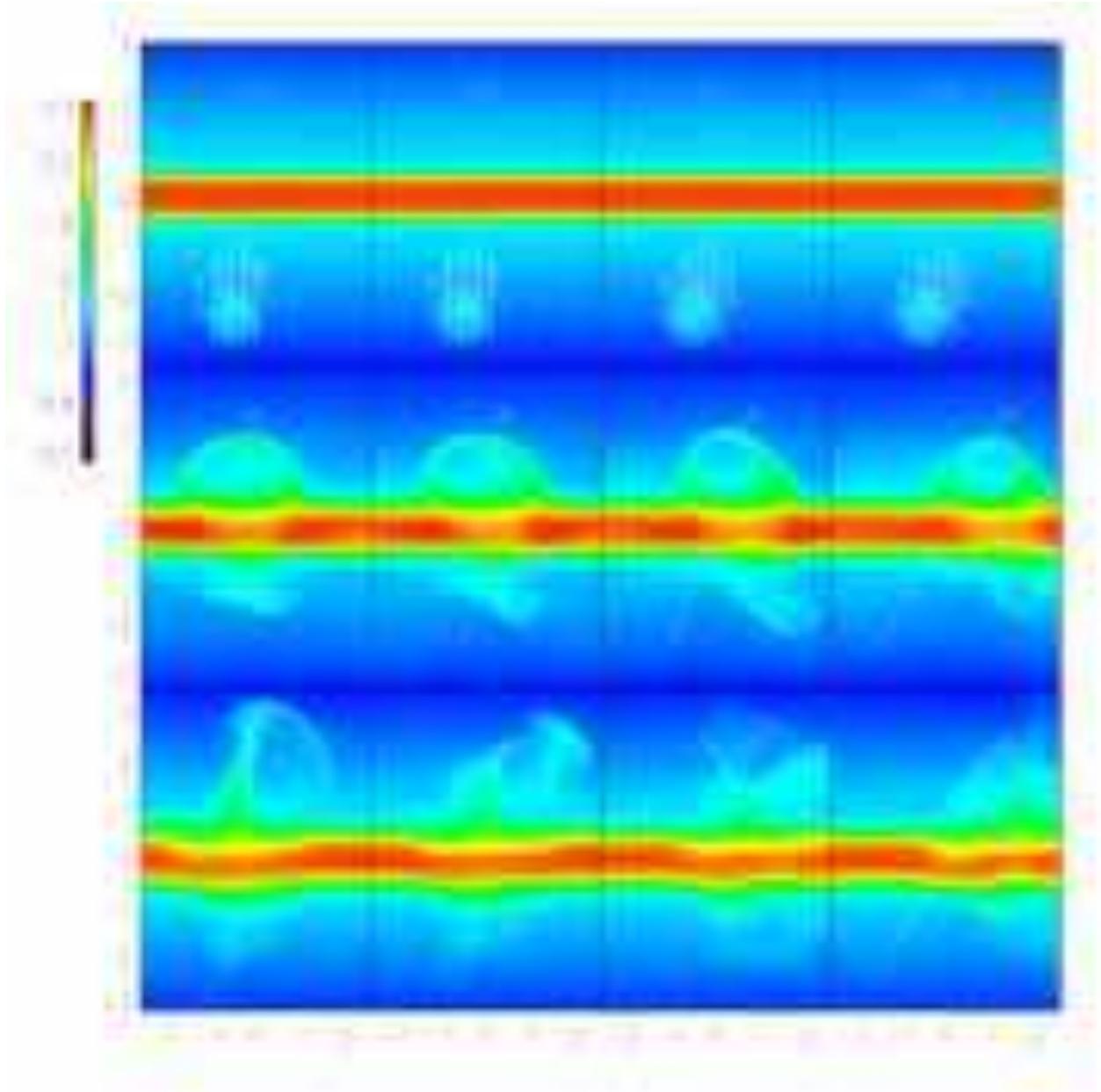}
\figcaption
{Column-density on the $x-z$ plane (integrating along the $y$-axis)
at three epochs ($t/t_0=0$ (top panels), $2.4$ (middle panels), and $4.8$ (bottom panels), respectively), 
for four models  O5 ($\theta_i =5 \arcdeg$), O10 ($10 \arcdeg$), O20 ($20 \arcdeg$), and O30 ($30 \arcdeg$).
At $t/t_0=0$ (top panels), initial velocity fields on the $y=0$ plane are shown. 
White arrows in the middle and bottom panels represent the initial trajectory of the cloud.
The image scales are logarithmic for $\sigma=\int \rho dl$.}
\end{figure}

\begin{figure}
\plotone{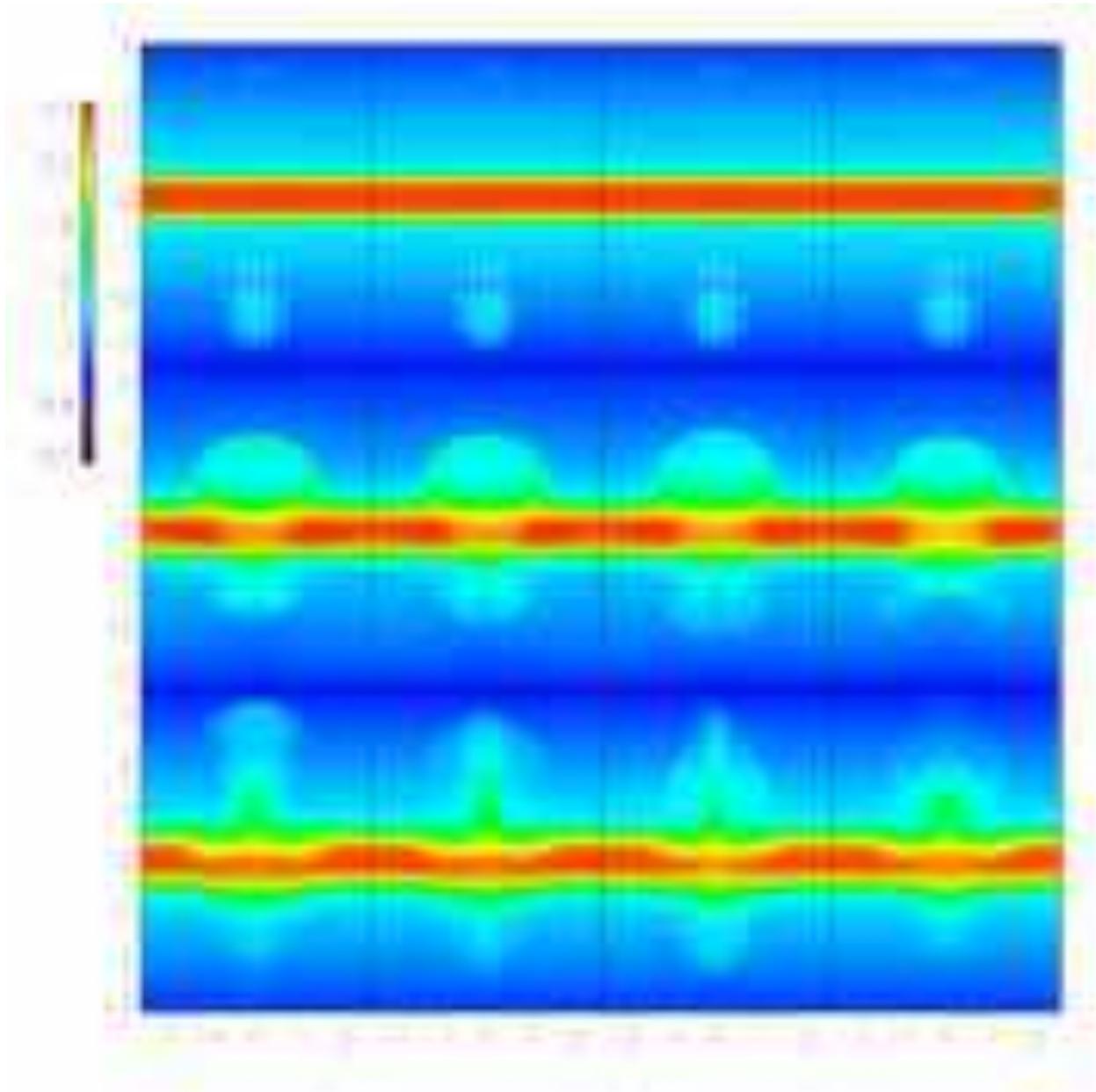}
\figcaption
{Same as Fig.\,6, but on the $y-z$ plane (integrating along the $x$-axis).}
\end{figure}

\begin{figure}
\plotone{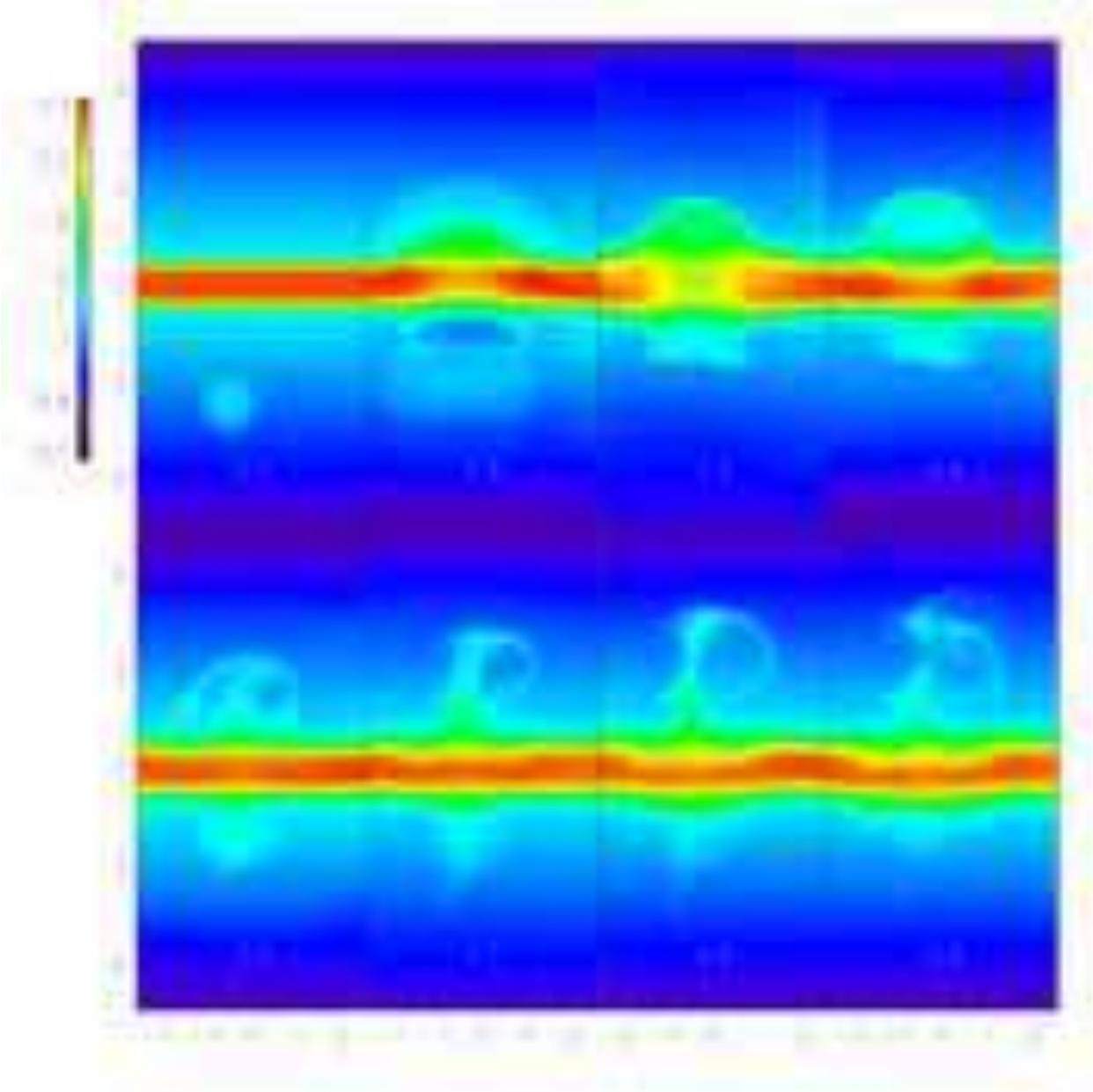}
\figcaption
{Time evolution of column-density of cool gas on the $x-z$ plane.
The snapshots are made at $t/t_0 = 0, 0.8, 1.6, 2.4$ (top:left to right),
$3.2, 4.0, 4.8$, and $5.6$ (bottom:left to right) for model O3.}
\end{figure}

\begin{figure}
\plotone{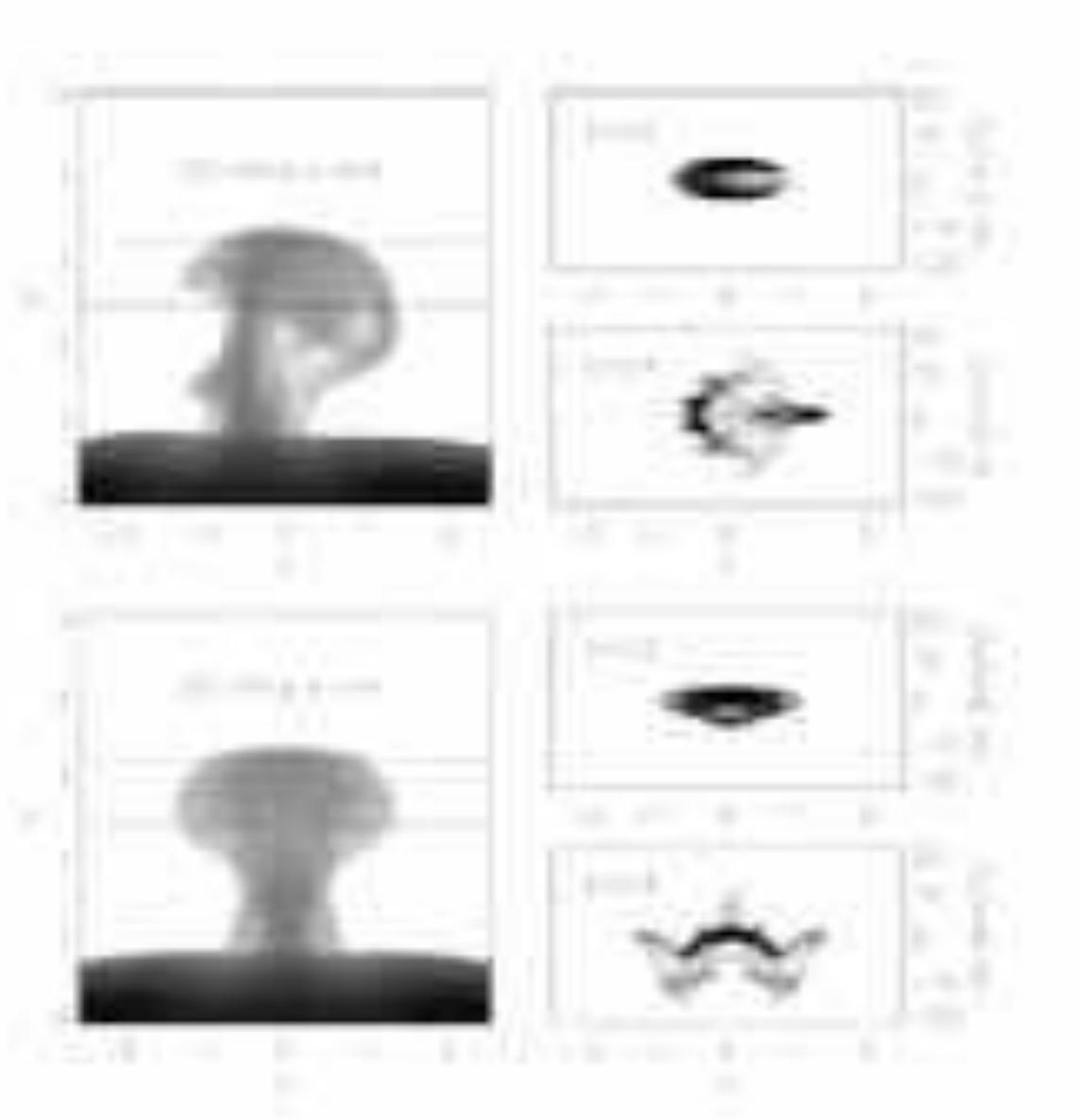}
\figcaption
{Column-density (left panels) and position-velocity (right panels) 
at $t/t_0=4.8$ in model O3.
Position-velocity is measured for the lower ($z=2.4 H_0$) and upper ($z=3.2 H_0$) regions
of the caps on (a) the $x-z$ plane (integrating along the $y$-axis) and on (b) the $y-z$ plane (integrating along the $x$-axis).}
\end{figure}

\end{document}